\documentclass[runningheads,envcountsect,envcountsame]{llncs}

\usepackage{graphicx}

\usepackage{hyperref}

\usepackage[capitalize]{cleveref}

\usepackage{xspace}
\usepackage[binary-units]{siunitx}

\usepackage[svgnames]{xcolor}
\colorlet{leepreparata}{orange}
\colorlet{makestep}{ForestGreen}
\colorlet{trapezoid}{blue}
\colorlet{delaunay}{red}

\newcommand{\Makestep}{\emph{Makestep}\xspace}
\newcommand{\Trapezoid}{\emph{Trapezoid}\xspace}
\newcommand{\Delaunay}{\emph{Delaunay}\xspace}
\newcommand{\LeePreparata}{\emph{Lee-Preparata}\xspace}

\newcommand{\etal}{\emph{et al.}}
\newcommand{\eps}{\varepsilon}

\usepackage{tikz}
\usetikzlibrary{calc}
\usetikzlibrary{backgrounds}
\usetikzlibrary{intersections}
\usetikzlibrary{patterns}
\usetikzlibrary{arrows.meta}
\usetikzlibrary{spy}
\tikzstyle{vertex}=[%
    draw,%
    circle,%
    minimum width=0.075cm,%
    fill,%
    outer sep=0,%
    inner sep=0,%
    ]
\tikzstyle{medianplot}=[%
    dashed,%
    ]
\tikzstyle{maxplot}=[%
    dotted,%
    opacity=1,%
    ]

\usepackage{pgfplots}
\usepackage{pgfplotstable}
\pgfplotsset{compat=1.14}
\pgfplotsset{every axis/.append style={%
    font=\small,%
    }}

\usepackage{subfig}

\usepackage{booktabs}
\usepackage[letterspace=50]{microtype}

\begin{document}
\title{An Experimental Study of Algorithms for
Geodesic Shortest Paths in the Constant-Workspace
Model\thanks{Supported in part by DFG projects MU/3501-1 and RO/2338-6 and
ERC StG 757609.}}
\titlerunning{Geodesic Shortest Path Algorithms with Constant Workspace}
\author{Jonas Cleve\orcidID{0000-0001-8480-1726} \and
Wolfgang Mulzer\orcidID{0000-0002-1948-5840}}
\authorrunning{J. Cleve and W. Mulzer}
\institute{Institut f\"ur Informatik, Freie Universit\"at Berlin,
Takustr.~9, 14195 Berlin, Germany
\email{[jonascleve, mulzer]@inf.fu-berlin.de}}
\maketitle              
\begin{abstract}
We perform an experimental evaluation of algorithms for
finding geodesic shortest paths between two points
inside a simple polygon in the
constant-workspace model.
In this model, the input resides in a read-only array that
can be accessed at random. In addition, the algorithm may
use a constant number of words for reading and for writing.
The constant-workspace model has been studied extensively
in recent years, and algorithms for geodesic shortest paths
have received particular attention.

We have implemented three such algorithms in Python, and we compare them
to the classic algorithm by Lee and Preparata that uses
linear time and linear space. We also clarify a few
implementation details that were missing in the original
description of the algorithms.  Our experiments show that
all algorithms perform as advertised in the original works and according to
the theoretical guarantees. However, the constant factors
in the running times turn out to be rather large for the
algorithms to be fully useful in practice.
\end{abstract}

\section{Introduction}%
\label{sec:introduction}

In recent years, the \emph{constant-workspace model} has
enjoyed growing popularity in the computational
geometry community~\cite{BanyassadyKoMu18}.
Motivated by the increasing deployment of small devices
with limited memory capacities, the goal is to develop
simple and efficient algorithms for the situation where
little workspace is available.
The model posits that the input resides in a read-only
array that can be accessed at random.
In addition, the algorithm may use a constant number
of memory words for reading and for writing.
The output must be written to a write-only memory
that cannot be accessed again for reading.
Following the initial work by Asano~\etal~from
2011~\cite{AsanoMuRoWa11b},
numerous results have been published for this model, leading
to a solid theoretical foundation for dealing with geometric
problems when the working memory is scarce. The recent survey by
Banyassady~\etal~\cite{BanyassadyKoMu18} gives an overview of the
problems that have been considered and of the
results that are available for them.

But how do these theoretical results measure up in practice,
particularly in view of the original motivation?
To investigate this question, we have implemented three
different constant-workspace algorithms for computing geodesic shortest
paths in simple polygons.
This is one of the first problems to be studied
in the constant-workspace model~\cite{AsanoMuRoWa11b,AsanoMuRoWa11a}.
Given that the general shortest path problem is unlikely
to be amenable to constant-workspace algorithms
(it is NL-complete~\cite{Tantau07}), it may come as a surprise
that a solution for the geodesic case exists at all.
By now, several algorithms are known, both for constant
workspace as well as in the \emph{time-space-trade-off} regime,
where the number of available cells of working memory may
range from constant to linear~\cite{Asano:Memoryconstrained_algorithms_simple:13,Har-Peled:Shortest_path_polygon:16}.

Due to the wide variety of approaches and the fundamental
nature of the problem, geodesic shortest paths are a
natural candidate for a deeper experimental study.
Our experiments show that all three constant-workspace algorithms work
well in practice and live up to their theoretical guarantees.
However, the large running times make them ill-suited for
very large input sizes.
During our implementation, we also noticed some missing
details in the original publications, and we explain below
how we have dealt with them.

As far as we know, our study constitutes the first large-scale
comparative evaluation of geometric algorithms in the
constant-workspace model. A previous implementation study, by
Baffier~\etal~\cite{BaffierDiKo18}, focused on time-space trade-offs
for stack-based algorithms and was centered on different applications 
of a powerful algorithmic
technique. Given the practical motivation and wide applicability of
constant-workspace algorithms for geometric problems, we hope that
our work will  lead to further experimental studies in this direction.

\section{The Four Shortest-Path Algorithms}%
\label{sec:algorithms}

We provide a brief summary for each of the four algorithms in our
implementation; further details can be found in the original
papers~\cite{AsanoMuRoWa11a,AsanoMuRoWa11b,LeePr84}. In each case, we use $P$ to denote
a simple input polygon in the plane with $n$ vertices.
We consider $P$ to be a closed, connected subset of the plane.
Given two points $s, t \in P$, our goal is to compute
a shortest path from $s$ to $t$ (with respect to the Euclidean
length) that lies completely inside $P$.

\subsection{The Classic Algorithm by Lee and Preparata}%
\label{sec:algorithms:lee-preparata}

\begin{figure}[t]
    \center%
    \includegraphics{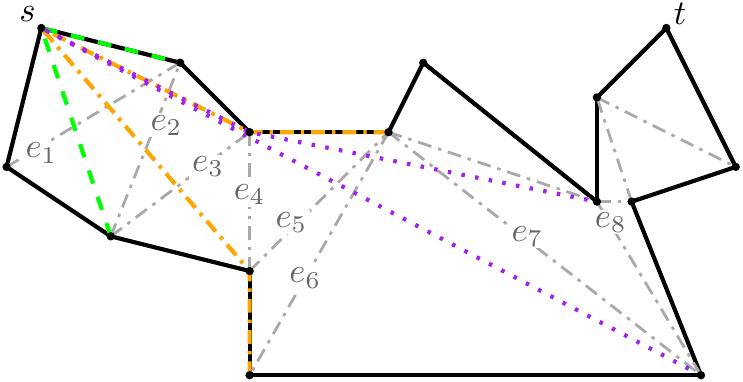}
    \caption{%
        Examples of three funnels during the algorithm for finding a
        shortest path from $s$ to $t$. Each has cusp $s$ and goes up to
        diagonals $e_2$ (green, dashed), $e_6$ (orange, dash
        dotted), and $e_8$ (purple, dotted).
    }\label{fig:lee-preparata-funnels}
\end{figure}

This is the classic linear-space algorithm for the geodesic shortest
path problem that can be found in
textbooks~\cite{LeePr84,Ghosh07}.
It works as follows: we triangulate $P$, and we find the triangle
that contains $s$ and the triangle that contains $t$.
Next, we determine the unique path between these two triangles
in the dual graph of the triangulation.
The path is unique since the dual graph of a triangulation of
a simple polygon is a tree~\cite{deBergChvKrOv08}.
We obtain a sequence $e_1, \dots, e_m$ of diagonals (incident to
pairs of consecutive triangles on the dual path) crossed by
the geodesic shortest path between $s$ and $t$, in that order.
The algorithm walks along these diagonals, while maintaining
a \emph{funnel}. The funnel consists of a \emph{cusp} $p$,
initialized to be $s$,
and two concave \emph{chains} from $p$ to the two endpoints of the
current diagonal $e_i$.
An example of these funnels can be found in
\cref{fig:lee-preparata-funnels}.
In each step $i$ of the algorithm, $i = 1, \dots, m-1$, we update
the funnel for $e_i$ to the funnel for $e_{i+1}$. There are two cases:
(i) if $e_{i+1}$ remains visible from the cusp $p$, we update the
appropriate concave chain, using a variant of Graham's scan;
(ii) if $e_{i+1}$ is no longer visible from $p$, we proceed along
the appropriate chain until we find the cusp for the next funnel.
We output the vertices encountered along the way as part
of the shortest path.
Implemented in the right way, this procedure takes linear time and
space.\footnote{If a triangulation of $P$ is already
available, the implementation is relatively straightforward.
If not, a linear-time implementation of the triangulation
procedure constitutes a significant challenge~\cite{Chazelle91}.
Simpler methods are available, albeit at the cost
of a slightly increased running time of
$O(n \log n)$~\cite{deBergChvKrOv08}.}

\subsection{Using Constrained Delaunay-Triangulations}%
\label{sec:algorithms:delaunay}

The first constant-workspace-algorithm for geodesic
shortest paths in simple polygons was presented by
Asano~\etal~\cite{AsanoMuRoWa11a}
in 2011. It is called \Delaunay, and it
constitutes a relatively direct adaptation
of the method of Lee and Preparata to the constant-workspace model.

\begin{figure}[t]
    \subfloat[The Delaunay triangulation of a point set.]{
        \includegraphics[scale=1,page=1]%
            {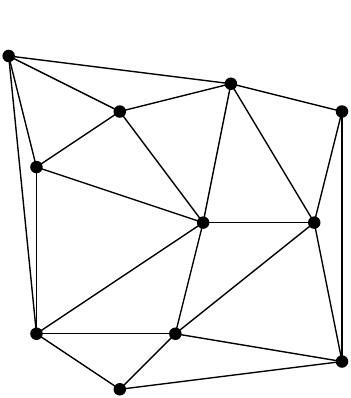}
    }
    \hfill
    \subfloat[
        A polygon on the point set which intersects the
        triangulation.]{
        \includegraphics[scale=1,page=2]%
            {figures/constrained-delaunay}
    }
    \hfill
    \subfloat[
        The circumcircle may contain points not visible from some
        triangle vertices.]{
        \includegraphics[scale=1,page=3]%
            {figures/constrained-delaunay}
    }
    \caption{%
        An example of a constrained Delaunay triangulation of a simple polygon.
    }\label{fig:constrained-delaunay-triangulation}
\end{figure}

In the constant-workspace model, we cannot explicitly compute
and store a triangulation of $P$. Instead, we use a uniquely defined
implicit triangulation of $P$, namely the \emph{constrained
Delaunay triangulation} of
$P$~\cite{Chew:Constrained_delaunay_triangulations:89}.
In this variant of the classic Delaunay triangulation, we prescribe
the edges of $P$ to be part of the desired triangulation.
Then, the additional triangulation edges cannot cross the prescribed edges.
Thus, unlike in the original Delaunay triangulation, the circumcircle
of a triangle may contain other vertices of $P$, as long as the line
segment from a triangle endpoint to the vertex crosses a prescribed
polygon edge, see \cref{fig:constrained-delaunay-triangulation} for an
example.

The constrained Delaunay triangulation of $P$ can
be navigated efficiently using constant workspace:
given a diagonal or a polygon edge, we can find the two incident
triangles in $O(n^2)$ time~\cite{AsanoMuRoWa11a}.
Using an $O(n)$ time constant-workspace-algorithm for finding
shortest paths in trees, also given by Asano~\etal~\cite{AsanoMuRoWa11a},
we can thus enumerate all triangles in the dual path between the constrained
Delaunay triangle that contains $s$ and the constrained
Delaunay triangle that contains $t$ in $O(n^3)$ time.

As in the algorithm by Lee and Preparata, we need to maintain
the visibility funnel while walking along the dual path of the 
constrained Delaunay triangulation.
Instead of the complete chains, we store only the two line segments
that define the current visibility cone (essentially the cusp together
with the first vertex of each chain). We recompute the two chains
whenever it becomes necessary.
The total running time of the algorithm is $O(n^3)$.
More details can be found in the paper by
Asano~\etal~\cite{AsanoMuRoWa11a}.

\subsection{Using Trapezoidal Decompositions}%
\label{sec:algorithms:trapezoid}

\begin{figure}[t]
    \hfill
    \subfloat[
        The trapezoidal decomposition is obtained by shooting rays up
        and down at every vertex.
        ]{
        \hspace{1.5em}
        \includegraphics[scale=1,page=1]%
            {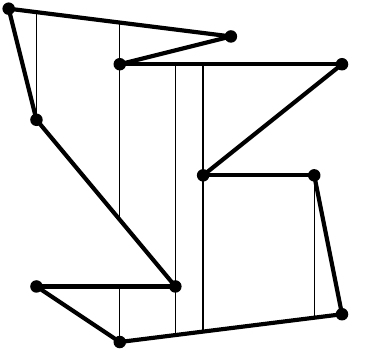}
        \hspace{1.5em}
    }
    \hfill
    \subfloat[
        Shifting all points to the right by $y\varepsilon$ makes sure
        no two share the same $x$-coordinate.
        ]{
        \hspace{1.5em}
        \includegraphics[scale=1,page=2]%
            {figures/trapezoid}
        \hspace{1.5em}
    }
    \hfill
    \null%
    \caption{%
        The trapezoidal decomposition of a polygon.
        If the polygon is in general position (right) each trapezoid
        has at most four neighbors which can all be found in $O(n)$
        time.
    }\label{fig:trapezoid-decomposition}
\end{figure}

This algorithm was also proposed by Asano~\etal~\cite{AsanoMuRoWa11a},
as a faster alternative to the
algorithm that uses constrained Delaunay triangulations.
It is based on the same principle as \Delaunay, but it uses the
trapezoidal decomposition of $P$ instead of the
Delaunay triangulation~\cite{deBergChvKrOv08}.
See \cref{fig:trapezoid-decomposition} for a depiction of the decomposition
and the symbolic perturbation method to avoid a general position assumption.
In the algorithm, we compute a trapezoidal decomposition of $P$, and we follow
the dual path between the trapezoid that contains $s$ and
the trapezoid that contains $t$, while maintaining a funnel
and outputting the new vertices of the geodesic shortest path
as they are discovered.
Assuming general position, we can find all incident
trapezoids of the current trapezoid and determine how to continue
on the way to $t$ in $O(n)$ time
(instead of $O(n^2)$ time in the case of the \Delaunay{} algorithm).
Since there are still $O(n)$ steps, the running time improves to $O(n^2)$.

\subsection{The Makestep Algorithm}%
\begin{figure}[t]
    \subfloat[
        $P'$ is the subset of $P$ cut off by the three points $p$,
        $q_1$, and $q_2$.
        Both points are convex, one is advanced.
    ]{
       \includegraphics[scale=1,page=1]%
            {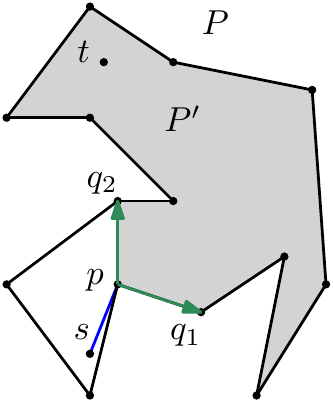}
    }
    \hfill
    \subfloat[
        Since $q_2$ is reflex we shoot a ray until it hits the
        boundary.
        $t$ lies to the left of this ray.
    ]{
        \includegraphics[scale=1,page=2]%
            {figures/makestep}
    }
    \hfill
    \subfloat[
        $p$ is relocated to the previous position of $q_2$ and $P'$ is
        shrunk along the ray.
    ]{
        \includegraphics[scale=1,page=3]%
            {figures/makestep}
    }
    \caption{%
        An illustration of the steps in the \Makestep{} algorithm.
    }\label{fig:makestep}
\end{figure}

\label{sec:algorithms:makestep}
This algorithm was presented by Asano~\etal~\cite{AsanoMuRoWa11b}.
It uses a direct approach to the geodesic shortest path problem and
unlike the two previous algorithms,
it does not try to mimic on the algorithm by Lee and Preparata. In
the traditional model, this approach would be deemed too
inefficient, but in the constant-workspace world, its simplicity
turns out to be beneficial. The main idea is as follows:
we maintain a \emph{current vertex} $p$ of the geodesic shortest path,
together with a \emph{visibility cone}, defined by two points $q_1$ and $q_2$
on the boundary of $P$.
The segments $pq_1$ and $pq_2$ cut off a subpolygon $P' \subseteq P$.
We maintain the invariant that the target $t$ lies in $P'$.
In each step, we gradually shrink $P'$ by advancing $q_1$ and $q_2$, sometimes
also relocating $p$ and outputting a new vertex of the geodesic shortest
path.
These steps are illustrated in \cref{fig:makestep}.
It is possible to realize the shrinking steps in such a
way that there are only $O(n)$ of them.
Each shrinking step takes $O(n)$ time, so the total running time of
the MakeStep algorithm is
$O(n^2)$.

\section{Our Implementation}%
\label{sec:implementation}

We have implemented the four algorithms from
Section~\ref{sec:algorithms} in
Python~\cite{PythonSoftwareFoundation:Python:}.
For graphical output and for plots, we use the
\texttt{matplotlib} library~\cite{Hunter:Matplotlib_2D_graphics:07}.
Even though there are some packages for Python that provide
geometric objects such as line segments, circles, etc.,
none of them seemed suitable for our needs.
Thus, we decided to implement all geometric primitives on our own.
The source code of the implementation is available online
in a Git-repository.\footnote{\url{https://github.com/jonasc/constant-workspace-algos}}

In order to apply the algorithm \LeePreparata, we must be
able to triangulate the simple input polygon $P$ efficiently.
Since implementing an efficient polygon triangulation algorithm
can be challenging and since this is not the main objective of our
study, we relied for this on the \emph{Python Triangle} library by
Rufat~\cite{Rufat:Python_Triangle:16}, a Python wrapper for
Shewchuk's \emph{Triangle}, which was written in
C~\cite{Shewchuk:Triangle_Engineering_2D:96}.
We note that \emph{Triangle} does not provide a linear-time
triangulation algorithm, which would be needed to achieve the
theoretically possible linear running time for the shortest path algorithm.
Instead, it contains three different implementations, namely
Fortune's sweep line algorithm, a randomized incremental
construction, and a divide-and-conquer method. All  three
implementations give a running time of $O(n\log{n})$.
For our study, we used the divide-and conquer algorithm, the default choice.
In the evaluation, we did not include the triangulation phase in the time and memory
measurement for running the algorithm by Lee and Preparata.

\subsection{General Implementation Details}

All three constant-workspace algorithms have been presented with a
general position assumption: \Delaunay{} and \Makestep{} assume that
no three vertices lie on a line, while
\Trapezoid{} assumes that no two vertices have the same
$x$-coordinate.
Our implementations  of \Delaunay{} and \Makestep{} also assume
general position, but they throw exceptions if a non-recoverable
general position violation is encountered.
Most violations, however, can be dealt with easily in our code; e.g.\@
when trying to find the constrained Delaunay triangle(s) for a
diagonal, we can simply ignore points collinear to this diagonal.
For the case of \Trapezoid,
Asano~\etal~\cite{AsanoMuRoWa11a}
described how to enforce the general position assumption by
changing the $x$-coordinate of every vertex to $x+\eps y$ for some
small enough $\eps > 0$ such that the $x$-order of all vertices is
maintained. In our implementation, we apply this method to every
polygon in which two vertices share the same $x$-coordinate.

The coordinates are stored as \SI{64}{\bit} IEEE 754 floats.
In order to prevent problems with floating point precision or
rounding, we take the following steps:
first, we never explicitly calculate angles, but we rely on the
usual three-point-orientation test, i.e., the computation of a
determinant to find the position of a point
$c$ relative to the directed line through to points $a$ and
$b$~\cite{deBergChvKrOv08}.
Second, if an algorithm needs to place a point somewhere in the
relative interior of a polygon edge, we store an
additional edge reference to account for inaccuracies when
calculating the new  point's coordinates.

\subsection{Implementing the Algorithm by Lee and Preparata}

The algorithm by Lee and Preparata can be implemented easily, in
a straightforward fashion.  There are no particular edge cases or
details that we need to take care of.
Disregarding the code for the geometric primitives, the algorithm
needs less then half as many lines of code than the other algorithms.

\subsection{Implementing Delaunay and Trapezoid}

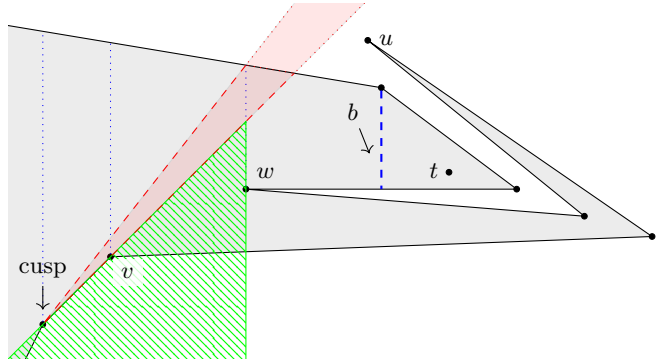
\begin{figure}[tb]
    \centering
    \begin{tikzpicture}[
            scale=0.9,
            trapezoid boundary/.style={blue, dotted},
            funnel boundary/.style={red, dashed},
        ]
        \path[clip] (-.5, -.5) rectangle (9.5, 4.75);
        \node[vertex] at (-.5, -1) (v0) {};
        \node[vertex] at (0, 0) (v1) {};
        \node[vertex] at (1, 1) (v2) {};
        \node[vertex] at (3, 2) (v3) {};
        \node[vertex] at (7, 2) (v4) {};
        \node[vertex] at (5, 3.5) (v5) {};
        \node[vertex] at (-1, 4.5) (v6) {};

        \node[vertex] at (9, 1.3) (e1) {};
        \node[vertex] at (4.8, 4.2) (e2) {};
        \node[vertex] at (8, 1.6) (e3) {};

        \begin{scope}[on background layer]
            \path[clip] (-.5, -.5) rectangle (9.5, 4.75);
            \draw[fill=lightgray!30!white, draw=black]
            ($(v0)$) -- ($(v1)$) -- ($(v2)$) -- ($(e1)$) --
            ($(e2)$) -- ($(e3)$) -- ($(v3)$) -- ($(v4)$) --
            ($(v5)$) -- ($(v6)$) -- ($(v0)$);
        \end{scope}
        \path[name path=top edge] (v5) -- (v6);
        \path[name path=v1 up] (v1) |- (v6);
        \path[name intersections={of=top edge and v1 up, name=i}]
        coordinate (t1) at (i-1);
        \path[name path=v2 up] (v2) |- (v6);
        \path[name intersections={of=top edge and v2 up, name=i}]
        coordinate (t2) at (i-1);
        \path[name path=v3 up] (v3) |- (v6);
        \path[name intersections={of=top edge and v3 up, name=i}]
        coordinate (t3) at (i-1);
        \path[name path=bot edge] (v3) -- (v4);
        \path[name path=v5 down] (v5) |- (v2);
        \path[name intersections={of=bot edge and v5 down, name=i}]
        coordinate (b5) at (i-1);
        \draw[trapezoid boundary] (v1) -- (t1);
        \draw[trapezoid boundary] (v2) -- (t2);
        \draw[trapezoid boundary,name path=boundary3] (v3) -- (t3);
        \draw[trapezoid boundary,dashed,thick] (v5) -- (b5);

        \path[name path=lower funnel] (v1) -- ($(v1)!5!(v2)$);
        \path[name intersections={of=boundary3 and lower funnel, name=i}]
        coordinate (c1) at (i-1);
        \draw[green, pattern color=green, pattern=north west lines]
        (-1,-1) -- (c1) -- (3,-1);

        \path[name intersections={of=top edge and lower funnel, name=i}]
        coordinate (fl) at (i-1);
        \draw[fill=red, opacity=0.1]
        ($(t3)$) -- ($(v1)$) -- ($(fl)$) -- cycle;
        \draw[funnel boundary] (t3) -- (v1) -- (fl);

        \path[name path=upper funnel] (v1) -- ($(v1)!5!(t3)$);
        \draw[dotted, red]
        ($(v1)!5!(t3)$) -- (t3) -- (fl) -- ($(v1)!5!(v2)$);
        \draw[fill=red, opacity=0.1]
        ($(v1)!5!(t3)$) -- (t3) -- (fl) -- ($(v1)!5!(v2)$);

        \node[anchor=south] at ([yshift=15pt]v1.north) (cusplabel) {cusp};
        \draw[->] (cusplabel) -- ([yshift=5pt]v1.north);

        \node[anchor=east] at ([xshift=-5pt,yshift=-10pt]v5.west)
        (edgelabel) {\(b\)};
        \draw[->] (edgelabel) -- ([yshift=15pt,xshift=-5pt]b5.north);

        \node[vertex] at (6,2.25) (t) {};
        \node[anchor=east] at (t.north west) {$t$};

        \node[anchor=south west] at (v3.north east) (v6label) {$w$};
        \node[anchor=west] at (e2.east) (v4label) {$u$};
        \node[anchor=north west,fill=white,white,opacity=0.8]
        at (v2.south east) {$v$};
        \node[anchor=north west] at (v2.south east) (v2label) {$v$};
    \end{tikzpicture}
    \caption{
        During the gift wrapping from the cusp to the diagonal $b$,
        the vertices need to be restricted to the shaded area.
        Otherwise, $u$ would be considered to be part of the
        geodesic shortest path, as it is to the left of $vw$.
    }%
    \label{fig:implementation:jarvis-march-x-constraint}
\end{figure}

In both constant-workspace adaptations of the algorithm by
Lee and Preparata, we encounter the following problem:
whenever the cusp of the current funnel changes, we need to
find the cusp of the new funnel, and we need to find the piece of the
geodesic shortest path that connects the former cusp to the
new cusp.
In their description of the algorithm,
Asano~\etal~\cite{AsanoMuRoWa11a} only say that this should be done
with an application of gift
wrapping (Jarvis' march)~\cite{deBergChvKrOv08}.
While implementing these two algorithms, we noticed that a naive
gift wrapping step that considers all the  vertices on $P$ between
the cusp of the current funnel and the next diagonal might include
vertices that are not visible inside the polygon.
\Cref{fig:implementation:jarvis-march-x-constraint} shows an example:
here $b$ is the next diagonal, and naively we would look at all
vertices along the polygon boundary between  $v$ and $w$.
Hence, $u$ would be considered as a gift wrapping candidate, and
since it forms the largest angle with the cusp and $v$
(in particular, an angle that is larger than the angle formed by
$w$) it would be chosen as the next point, even though $w$ should
be the cusp of the next funnel.
A simple fix for this problem would be an explicit check for
visibility in each gift-wrapping step. Unfortunately,
the resulting increase in the running time would be too expensive 
for a realistic implementation of the algorithms.

Our solution for \Trapezoid{} is to consider only vertices whose
$x$-coordinate is between the cusp of the current vertex and the
point where the current visibility cone crosses the boundary
of $P$ for the first time.
For ease of implementation, one can also limit it to the
$x$-coordinate of the last trapezoid boundary visible from the cusp.
\Cref{fig:implementation:jarvis-march-x-constraint} shows this as
the dotted green region.
For \Delaunay, a similar approach can be used.
The only difference is that the triangle boundaries in general
are not vertical lines.

\subsection{Implementing Makestep}

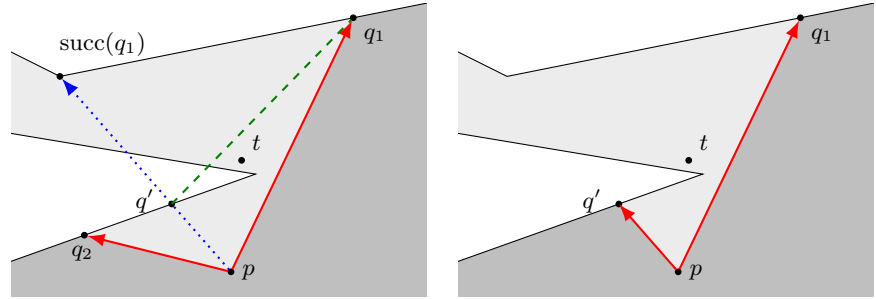
\begin{figure}[tb]
    \hfill
    \begin{tikzpicture}[
            xscale=0.65,
            yscale=0.65,
            ray shot/.style={blue, dotted, thick, -Latex},
            divider/.style={green!50!black, dashed, thick},
            funnel boundary/.style={red, thick, -Latex},
        ]
        \path[clip] (0, 1.5) rectangle (8.5, 7.5);

        \node[vertex] at (4.5, 2) (p) {};

        \node[vertex] at (-2, 1.5) (v1) {};
        \node[] at (5, 4) (v2) {};
        \node[vertex] at (-1, 5) (v3) {};

        \node[vertex] at (11, 8) (v4) {};
        \node[vertex] at (1, 6) (v5) {};
        \node[vertex] at (-1, 7) (v6) {};

        \node[vertex] at ($(v4)!.4!(v5)$) (q1) {};
        \node[vertex] at ($(v1)!.5!(v2)$) (q2) {};
        \node[anchor=north west] at (q1.south east) {$q_1$};
        \node[anchor=north] at (q2.south west) {$q_2$};
        \node[anchor=south west] at ([yshift=5pt,xshift=-5pt]v5.north)
        {$\text{succ}(q_1)$};
        \node[anchor=west] at (p.east) {$p$};

        \draw[funnel boundary] (p) -- (q1);
        \draw[funnel boundary] (p) -- (q2);

        \path[name path=boundary] (v1) -- (v2);
        \draw[ray shot, name path=succ ray] (p) -- (v5);
        \path[vertex, name intersections={of=boundary and succ ray, name=i}]
        coordinate (qprime) at (i-1);
        \node[vertex] at (qprime) {};
        \node[anchor=east] at ([xshift=-5pt,yshift=2pt]qprime.west)
        {$q'$};

        \draw[divider] (qprime) -- (q1);

        \coordinate (t on line) at ($(qprime)!(v2)!(q1)$);
        \node[vertex] at ($(t on line)!.5!(v2)$) (t) {};
        \node[anchor=south west] at (t.north east) {$t$};

        \begin{scope}[on background layer]
            \path[clip] (0, 1.5) rectangle (8.5, 7.5);
            \fill[fill=lightgray!30!white]
            (-1, -1) -- (10, -1) -- ($(v4)$) --
            ($(v5)$) -- ($(v6)$) -- ($(v3)$) --
            ($(v2)$) -- ($(v1)$) -- cycle;
            \fill[fill=lightgray]
            (-1, -1) -- (10, -1) -- ($(v4)$) --
            ($(q1)$) -- ($(p)$) -- ($(q2)$) --
            ($(v1)$) -- cycle;
            \draw[draw=black]
            (-1, -1) -- (10, -1) -- ($(v4)$) --
            ($(v5)$) -- ($(v6)$) -- ($(v3)$) --
            ($(v2)$) -- ($(v1)$) -- cycle;
        \end{scope}
    \end{tikzpicture}
    \hfill
    \begin{tikzpicture}[
            xscale=0.65,
            yscale=0.65,
            ray shot/.style={blue, dotted, thick, -Latex},
            divider/.style={green!50!black, dashed, thick},
            funnel boundary/.style={red, thick, -Latex},
        ]
        \path[clip] (0, 1.5) rectangle (8.5, 7.5);

        \node[vertex] at (4.5, 2) (p) {};

        \node[vertex] at (-2, 1.5) (v1) {};
        \node[] at (5, 4) (v2) {};
        \node[vertex] at (-1, 5) (v3) {};

        \node[vertex] at (11, 8) (v4) {};
        \node[] at (1, 6) (v5) {};
        \node[vertex] at (-1, 7) (v6) {};

        \node[vertex] at ($(v4)!.4!(v5)$) (q1) {};
        \node[anchor=north west] at (q1.south east) {$q_1$};
        \node[anchor=west] at (p.east) {$p$};

        \path[name path=boundary] (v1) -- (v2);
        \path[vertex, name intersections={of=boundary and succ ray, name=i}]
        coordinate (qprime) at (i-1);
        \node[vertex] at (qprime) {};
        \node[anchor=east] at ([xshift=-5pt,yshift=2pt]qprime.west)
        {$q'$};

        \draw[funnel boundary] (p) -- (q1);
        \draw[funnel boundary] (p) -- (qprime);

        \coordinate (t on line) at ($(qprime)!(v2)!(q1)$);
        \node[vertex] at ($(t on line)!.5!(v2)$) (t) {};
        \node[anchor=south west] at (t.north east) {$t$};

        \begin{scope}[on background layer]
            \path[clip] (0, 1.5) rectangle (8.5, 7.5);
            \fill[fill=lightgray!30!white]
            (-1, -1) -- (10, -1) -- ($(v4)$) --
            ($(v5)$) -- ($(v6)$) -- ($(v3)$) --
            ($(v2)$) -- ($(v1)$) -- cycle;
            \fill[fill=lightgray]
            (-1, -1) -- (10, -1) -- ($(v4)$) --
            ($(q1)$) -- ($(p)$) -- ($(qprime)$) --
            ($(v1)$) -- cycle;
            \draw[draw=black]
            (-1, -1) -- (10, -1) -- ($(v4)$) --
            ($(v5)$) -- ($(v6)$) -- ($(v3)$) --
            ($(v2)$) -- ($(v1)$) -- cycle;
        \end{scope}
    \end{tikzpicture}
    \hfill\null%
    \caption{
        Left: Asano~\etal~\cite{AsanoMuRoWa11b} state
        that one should check whether ``$t$ lies in the
        subpolygon from $q'$ to $q_1$.''
        This subpolygon, however, is not clearly defined as the
        line segment $q'q_1$ does not lie inside $P$.
        Considering $pq'$ instead and using $q_1pq'$ to shrink
        the cutoff region gives the correct result on the right.
    }%
    \label{fig:implementation:makestep-wrong-decision}
\end{figure}

Our implementation of the \Makestep{} algorithm is also relatively
straightforward. Nonetheless, we would
like to point out one interesting detail;  see \cref{fig:implementation:makestep-wrong-decision}.
The description by Asano~\etal~\cite{AsanoMuRoWa11b} says that
to advance the visibility cone, we should check if
``\emph{$t$ lies in the subpolygon from $q'$ to $q_1$}.''
If so, the visibility cone should be shrunk to $q'pq_1$, otherwise
to $q_2pq'$.

However, the ``\emph{subpolygon from $q'$ to $q_1$}''
is not clearly defined for the case that the line segment
$q'q_1$ is not contained in $P$.
To avoid this difficulty, we instead consider the line segment $pq'$.
This line segment is always contained in $P$, and it divides the
cutoff region $P'$ into two parts, a ``subpolygon''
between $q'$ and $q_1$ and a ``subpolygon'' between $q_2$ and $q'$.
Now we can easily choose the one containing $t$.

\section{Experimental Setup}%
\label{sec:experiments}

We now describe how we conducted the experimental evaluation
of our four implementations for geodesic shortest path algorithms.

\subsection{Generating the Test Instances}

Our experimental approach is as follows:
given a desired number of vertices $n$, we generate
4--10 (pseudo)random polygons with $n$ vertices.
For this, we use a tool developed in a software project carried 
out under the supervision of G\"unter Rote at
the Institute of Computer Science at Freie Universit\"at
Berlin~\cite{Dierker12}. Among others, the tool  provides an
implementation of the \emph{Space Partitioning} algorithm for
generating random simple polygons presented by Auer and
Held~\cite{AuerHe96}.

Next, we generate the set $S$ of desired endpoints for the geodesic
shortest paths.
This is done as follows: for each edge $e$ of each generated
polygon, we find the incident triangle $t_e$ of $e$ in
the constrained Delaunay triangulation of the polygon.
We add the barycenter of $t_e$ to $S$. In the end, the set
$S$ will have between $\left\lfloor n/2 \right\rfloor$ and $n-2$
points. We will compute the geodesic shortest path for each
pair of distinct points in $S$.

\subsection{Executing the Tests}

For each pair of points $s, t \in S$, we find the
geodesic shortest path between $s$ and $t$ using each of the
four implemented algorithms.
Since the number of pairs grows quadratically in $n$, we restrict
the tests to $1500$ random pairs for all $n\geq200$.

First, we run each algorithm once in order to assess the memory
consumption.
This is done by using the \texttt{get\_traced\_memory} function of
the built-in \texttt{tracemalloc} module which returns the peak and
current memory consumption---the difference tells us how much memory
was used by the algorithm.
Starting the memory tracing just before running the algorithm gives the
correct values for the peak memory consumption.
In order to obtain reproducible numbers we also disable Python's
garbage collection functionality using the built-in
\texttt{gc.disable} and \texttt{gc.enable} functions.

After that, we run the algorithm between $5$ and $20$ times,
depending on how long it takes.
We measure the processor time for each run with the
\texttt{process\_time} function of the \texttt{time} module which gives
the time during which the process was active on the processor in user
and in system mode.
We then take the median of the times as a representative running time
for this point pair.

\subsection{Test Environment}

Since we have a quadratic number of test cases for each instance,
our experiments take a lot of time.
Thus, the tests were distributed on multiple machines and on
multiple cores.  We had six computing machines at our disposal,
each with two quad-core \textls*{CPU}s.
Three machines had Intel Xeon E5430 \textls*{CPU}s with \SI{2.67}{\GHz}; the
other three had \textls*{AMD} Opteron 2376 \textls*{CPU}s with \SI{2.3}{\GHz}.
All machines had \SI{32}{\giga\byte} RAM, even though, as can be seen
in the next section, memory was never an issue.
The operating system was a Debian 8 and we used version 3.5 of the
Python interpreter to implement the algorithms and to execute the tests.

\section{Experimental Results}%
\label{sec:results}

\begin{figure}[tb]
    \centering
    \begin{tikzpicture}[
            spy using outlines={
                    rectangle,
                    connect spies,
                },
        ]
        \begin{axis}[
                width=1.1\linewidth,
                height=0.8\linewidth,
                title={\# vertices vs.\ memory in \si{\kilo\byte}},
                legend style={
                        at={(0.1, 0.96)},
                        anchor=north west,
                    },
                legend cell align=left,
                legend columns=2,
                scaled y ticks={base 10:-3},
                ytick scale label code/.code={},
                yticklabel style={anchor=west,outer sep=2pt,fill=white},
                enlargelimits={rel=.06},
                ymin=0.35,
                grid=major,
                grid style={dotted},
                /pgf/number format/1000 sep={},
                title style={anchor=center},
            ]
            \addplot[
                leepreparata,
                mark=triangle,
                only marks,
            ] table[x=n,y expr={\thisrow{lee_preparata}}]
                {figures/random-memory.table};
            \addlegendentry{\LeePreparata{} $O(n)$}
            \addplot[
                makestep,
                mark=o,
                only marks,
            ] table[x=n,y expr={\thisrow{makestep}}]
                {figures/random-memory.table};
            \addlegendentry{\Makestep{} $O(1)$}
            \addplot[
                trapezoid,
                mark=diamond,
                only marks
            ] table[x=n,y expr={\thisrow{trapezoid}}]
                {figures/random-memory.table};
            \addlegendentry{\Trapezoid{} $O(1)$}
            \addplot[
                delaunay,
                mark=square,mark size=1.5pt,
                only marks
            ] table[x=n,y expr={\thisrow{delaunay}}]
                {figures/random-memory.table};
            \addlegendentry{\Delaunay{} $O(1)$}

            \addplot[
                leepreparata,
                opacity=0.75,
                mark=x,
                only marks,
            ] table[x=n,y expr={\thisrow{lee_preparata}}]
                {figures/random-memory.max.table};
            \addplot[
                makestep,
                opacity=0.75,
                mark=x,
                only marks,
            ] table[x=n,y expr={\thisrow{makestep}}]
                {figures/random-memory.max.table};
            \addplot[
                trapezoid,
                opacity=0.75,
                mark=x,
                only marks
            ] table[x=n,y expr={\thisrow{trapezoid}}]
                {figures/random-memory.max.table};
            \addplot[
                delaunay,
                opacity=0.75,
                mark=x,
                only marks
            ] table[x=n,y expr={\thisrow{delaunay}}]
                {figures/random-memory.max.table};
            \input{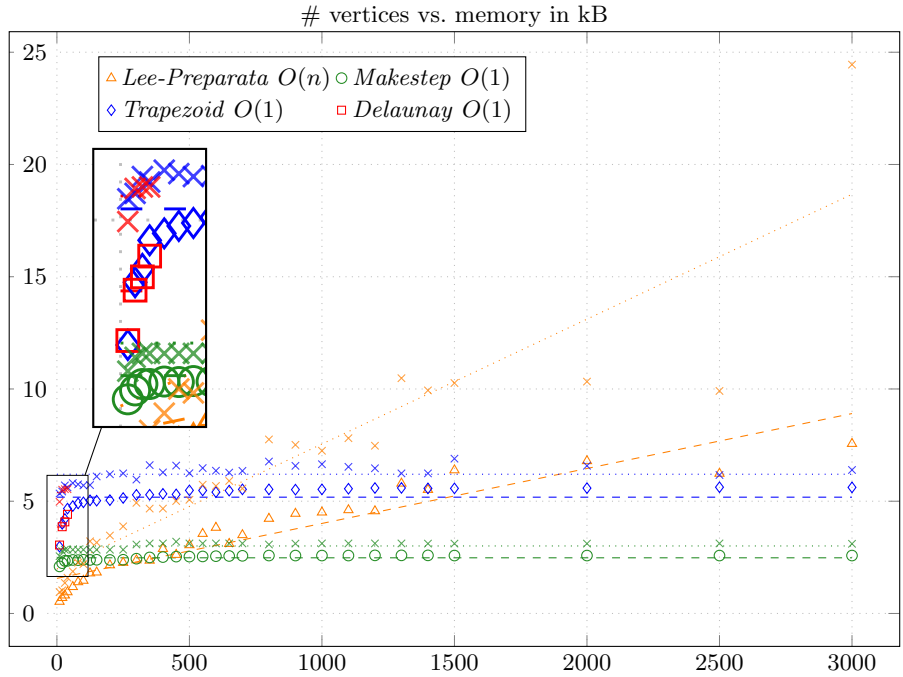}
            \coordinate (spycenter) at (axis cs:40,3900);
            \coordinate (glass) at (axis cs:350,14500);
            \spy[magnification=2.75,size=1.5cm,height=3.7cm] on (spycenter) in node[fill=white] at (glass);
        \end{axis}
    \end{tikzpicture}
    \caption{
        Memory consumption for random instances.
        The outlined shapes are the median values; the semi-transparent crosses are maximum values.
    }%
    \label{fig:random-memory-plot}
\end{figure}

\begin{figure}[p]
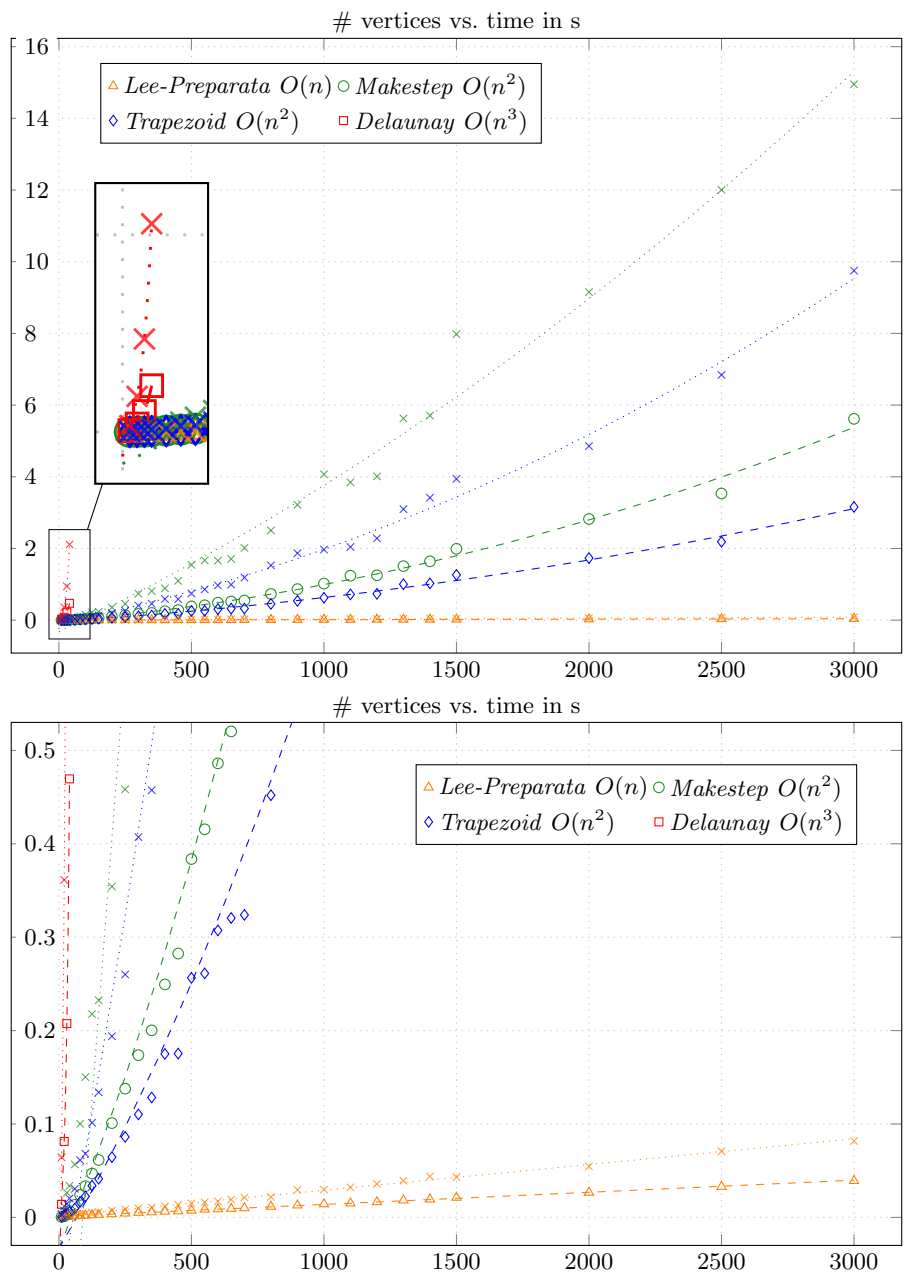

    \centering
    \begin{tikzpicture}[
        spy using outlines={
                rectangle,
                connect spies,
            },
        ]
        \begin{axis}[
                width=1.1\linewidth,
                height=0.8\linewidth,
                title={\# vertices vs.\ time in \si{\second}},
                legend style={
                        at={(0.1, 0.96)},
                        anchor=north west,
                    },
                legend columns=2,
                legend cell align=left,
                yticklabel style={anchor=west,outer sep=2pt,fill=white},
                enlargelimits={rel=.06},
                ymin=0,
                grid=major,
                grid style={dotted},
                /pgf/number format/1000 sep={},
                title style={anchor=center},
            ]
            \addplot[
                leepreparata,
                mark=triangle,
                only marks,
            ] table[x=n,y=lee_preparata] {figures/random-time.table};
            \addlegendentry{\LeePreparata{} $O(n)$}
            \addplot[
                makestep,
                mark=o,
                only marks,
            ] table[x=n,y=makestep] {figures/random-time.table};
            \addlegendentry{\Makestep{} $O(n^2)$}
            \addplot[
                trapezoid,
                mark=diamond,
                only marks
            ] table[x=n,y=trapezoid] {figures/random-time.table};
            \addlegendentry{\Trapezoid{} $O(n^2)$}
            \addplot[
                delaunay,
                mark=square,mark size=1.5pt,
                only marks
            ] table[x=n,y=delaunay] {figures/random-time.table};
            \addlegendentry{\Delaunay{} $O(n^3)$}

            \addplot[
                leepreparata,
                opacity=0.75,
                mark=x,
                only marks,
            ] table[x=n,y=lee_preparata]
                {figures/random-time.max.table};
            \addplot[
                makestep,
                opacity=0.75,
                mark=x,
                only marks,
            ] table[x=n,y=makestep] {figures/random-time.max.table};
            \addplot[
                trapezoid,
                opacity=0.75,
                mark=x,
                only marks
            ] table[x=n,y=trapezoid] {figures/random-time.max.table};
            \addplot[
                delaunay,
                opacity=0.75,
                mark=x,
                only marks
            ] table[x=n,y=delaunay] {figures/random-time.max.table};
            \input{figures/random-time}
            \coordinate (spycenter) at (axis cs:40,1);
            \coordinate (glass) at (axis cs:350,8);
            \spy[magnification=2.75,size=1.5cm,height=4cm] on (spycenter) in node[fill=white] at (glass);
        \end{axis}
    \end{tikzpicture}

    \begin{tikzpicture}
        \begin{axis}[
                width=1.1\linewidth,
                height=0.7\linewidth,
                title={\# vertices vs.\ time in \si{\second}},
                legend style={
                        at={(0.95, 0.92)},
                        anchor=north east,
                    },
                legend columns=2,
                legend cell align=left,
                yticklabel style={anchor=west,outer sep=2pt,fill=white},
                enlargelimits={rel=.06},
                ymin=0,
                ymax=.5,
                grid=major,
                grid style={dotted},
                /pgf/number format/1000 sep={},
                title style={anchor=center},
            ]
            \addplot[
                leepreparata,
                mark=triangle,
                only marks,
            ] table[x=n,y=lee_preparata] {figures/random-time.table};
            \addlegendentry{\LeePreparata{} $O(n)$}
            \addplot[
                makestep,
                mark=o,
                only marks,
            ] table[x=n,y=makestep] {figures/random-time.table};
            \addlegendentry{\Makestep{} $O(n^2)$}
            \addplot[
                trapezoid,
                mark=diamond,
                only marks
            ] table[x=n,y=trapezoid] {figures/random-time.table};
            \addlegendentry{\Trapezoid{} $O(n^2)$}
            \addplot[
                delaunay,
                mark=square,mark size=1.5pt,
                only marks
            ] table[x=n,y=delaunay] {figures/random-time.table};
            \addlegendentry{\Delaunay{} $O(n^3)$}

            \addplot[
                leepreparata,
                opacity=0.75,
                mark=x,
                only marks,
            ] table[x=n,y=lee_preparata]
                {figures/random-time.max.table};
            \addplot[
                makestep,
                opacity=0.75,
                mark=x,
                only marks,
            ] table[x=n,y=makestep] {figures/random-time.max.table};
            \addplot[
                trapezoid,
                opacity=0.75,
                mark=x,
                only marks
            ] table[x=n,y=trapezoid] {figures/random-time.max.table};
            \addplot[
                delaunay,
                opacity=0.75,
                mark=x,
                only marks
            ] table[x=n,y=delaunay] {figures/random-time.max.table};
            \input{figures/random-time}
        \end{axis}
    \end{tikzpicture}
    \caption{
        Runtime for random instances.
        Outlined shapes are median values; semi-transparent crosses are
        maximum values.
        The bottom plot is a scaled version of the top.
    }%
    \label{fig:random-time-plot}
\end{figure}

The results of the experiments can be seen in the following plots.
The plot in \Cref{fig:random-memory-plot} shows the median and
maximum memory consumption as solid shapes and transparent crosses,
respectively, for each algorithm and for each input size.
More precisely, the plot shows the median and the maximum over all
polygons with a given size and over all pairs of points in each
such polygon.

We observe that the memory consumption for \Trapezoid{} and for
\Makestep{} is always smaller than a certain constant.
At first glance, the shape of the median values might suggest
logarithmic growth. However, a smaller number of vertices leads
to a higher probability that $s$ and $t$ are directly visible to
each other.
In this case, many geometric functions and subroutines, each of
which requires an additional constant amount of memory, are not called.
A large number of point pairs with only small memory consumption
naturally entails a smaller median value.
We can observe a very similar effect in the memory
consumption of the \LeePreparata{} algorithm for small values
of $n$. However, as $n$ grows, we can see that the memory requirement
begins to grow linearly with $n$.

The second plot in \Cref{fig:random-time-plot} shows the median
and the maximum running time in the same way as \Cref{fig:random-memory-plot}.
Not only does \Delaunay{} have a cubic running time, but it also
seems to exhibit a quite large constant: it grows much faster than
the other algorithms.

In the lower part of \Cref{fig:random-time-plot}, we see the same
$x$-domain, but with a much smaller $y$-domain.
Here, we observe that \Trapezoid{} and \Makestep{} both have a quadratic
running time; \Trapezoid{} needs about two thirds of the time required
by \Makestep. Finally, the linear-time behavior of \LeePreparata{} can
clearly be discerned.

Additionally, we observed that the tests ran approximately
\SI{85}{\percent} slower on the AMD machines than on the Intel servers.
This reflects the difference between the clock speeds of 
\SI{2.3}{\GHz} and \SI{2.67}{\GHz}.
Since the tests were distributed equally on the machines, this does
not change the overall qualitative results and the comparison between
the algorithms.

\section{Conclusion}%
\label{sec:conclusion}

We have implemented and experimented with three different
constant-workspace algorithms for geodesic shortest paths in simple
polygons.  Not only did we observe the cubic worst-case running time
of \Delaunay, but we also noticed that the constant factor is rather
large.  This renders the algorithm virtually useless already for
polygons with a few hundred vertices, where the shortest path
computation might, in the worst case, take several minutes.

As predicted by the theory, \Makestep{} and \Trapezoid{} exhibit the
same asymptotic running time and space consumption.
\Trapezoid{} has an advantage in the constant factor of the running
time, while \Makestep{} needs only about half as much memory.
Since in both cases the memory requirement is bounded by a
constant, \Trapezoid{} would be our preferred algorithm.

We chose Python for the implementation mostly due to our previous 
programming experience,
good debugging facilities, fast prototyping possibilities, and
the availability of numerous libraries.
In hindsight, it might have been better to choose another
programming language that allows for more low-level control of the 
underlying hardware. Python's memory profiling and tracking
abilities are limited, so that we cannot easily get a detailed view
of the used memory with all the variables.
Furthermore, a more detailed control of the memory management
could be useful for performing more detailed experiments.



\bibliographystyle{splncs04}
\bibliography{bibliography}

\appendix
\section{Tables of Experimental Results}
Here we list the experimental results shown in
\cref{fig:random-memory-plot,fig:random-time-plot}.
\sisetup{%
    ,table-format=5.0%
    ,table-auto-round%
    ,table-number-alignment=center%
}
\begin{table}
    \caption{
        The median and maximum memory usage in bytes for all runs with
        a specific number of vertices $n$.
    }
    \center%
    \begin{tabular}{S[table-format=4.0]SSSSSSSS}\toprule
{} & \multicolumn{2}{c}{\Delaunay} & \multicolumn{2}{c}{\LeePreparata} & \multicolumn{2}{c}{\Makestep} & \multicolumn{2}{c}{\Trapezoid}\\ \cmidrule(rl){2-3} \cmidrule(rl){4-5} \cmidrule(rl){6-7} \cmidrule(rl){8-9}
{$n$} & {median} & {max} & {median} & {max} & {median} & {max} & {median} & {max}\\
\midrule
10 & 3048 & 4976 & 528 & 952 & 2096 & 2552 & 2976 & 5344
 \\
20 & 3864 & 5512 & 696 & 1032 & 2240 & 2776 & 3992 & 5432
 \\
30 & 4080 & 5536 & 808 & 1360 & 2344 & 2840 & 4208 & 5704
 \\
40 & 4416.0 & 5536 & 952.0 & 1592 & 2344.0 & 2840 & 4672.0 & 5616
 \\[0.5em]
60 &  &  & 1184.0 & 1872 & 2384.0 & 2840 & 4784.0 & 5808
 \\
80 &  &  & 1400.0 & 2264 & 2376.0 & 2840 & 4904.0 & 5752
 \\
100 &  &  & 1464 & 2200 & 2392 & 2840 & 4952 & 5704
 \\
125 &  &  & 1792 & 3216 & 2384 & 2840 & 5040 & 5720
 \\[0.5em]
150 &  &  & 1832.0 & 3160 & 2392.0 & 2840 & 5024.0 & 6104
 \\
200 &  &  & 2152 & 3472 & 2384 & 2840 & 5048 & 6200
 \\
250 &  &  & 2264.0 & 3880 & 2392.0 & 2840 & 5144.0 & 6240
 \\
300 &  &  & 2376.0 & 4928 & 2440.0 & 3072 & 5284.0 & 5964
 \\[0.5em]
350 &  &  & 2360.0 & 4672 & 2496.0 & 3120 & 5288.0 & 6608
 \\
400 &  &  & 2880.0 & 4672 & 2512.0 & 3120 & 5328 & 6284
 \\
450 &  &  & 2616 & 5008 & 2532 & 3200 & 5304 & 6588
 \\
500 &  &  & 3048 & 5064 & 2532 & 3120 & 5484 & 6248
 \\[0.5em]
550 &  &  & 3552.0 & 5736 & 2540.0 & 3120 & 5480.0 & 6476
 \\
600 &  &  & 3824.0 & 5680 & 2552.0 & 3120 & 5404.0 & 6360
 \\
650 &  &  & 3104 & 5904 & 2560 & 3120 & 5472 & 6276
 \\
700 &  &  & 3496 & 5568 & 2568 & 3120 & 5528 & 6352
 \\[0.5em]
800 &  &  & 4224.0 & 7752 & 2580.0 & 3072 & 5528.0 & 6768
 \\
900 &  &  & 4448.0 & 7512 & 2580.0 & 3112 & 5516.0 & 6576
 \\
1000 &  &  & 4504.0 & 7248 & 2580.0 & 3120 & 5528.0 & 6648
 \\
1100 &  &  & 4608.0 & 7808 & 2580.0 & 3120 & 5556.0 & 6532
 \\[0.5em]
1200 &  &  & 4560.0 & 7472 & 2588.0 & 3120 & 5588.0 & 6468
 \\
1300 &  &  & 5792.0 & 10480 & 2588 & 3120 & 5592 & 6240
 \\
1400 &  &  & 5512.0 & 9936 & 2588.0 & 3112 & 5572 & 6240
 \\
1500 &  &  & 6384.0 & 10264 & 2580.0 & 3112 & 5572 & 6896
 \\[0.5em]
2000 &  &  & 6792.0 & 10328 & 2580.0 & 3112 & 5584.0 & 6580
 \\
2500 &  &  & 6232.0 & 9912 & 2580.0 & 3120 & 5624.0 & 6168
 \\
3000 &  &  & 7560.0 & 24448 & 2580.0 & 3104 & 5616.0 & 6392
 \\
\bottomrule\end{tabular}

\end{table}
\sisetup{table-format=2.6}
\begin{table}
    \caption{
        The median and maximum running times in seconds for all runs
        with a specific number of vertices $n$.}
    \center%
    \scriptsize
    \begin{tabular}{S[table-format=4.0]SSSSSSSS}\toprule
{} & \multicolumn{2}{c}{\Delaunay} & \multicolumn{2}{c}{\LeePreparata} & \multicolumn{2}{c}{\Makestep} & \multicolumn{2}{c}{\Trapezoid}\\ \cmidrule(rl){2-3} \cmidrule(rl){4-5} \cmidrule(rl){6-7} \cmidrule(rl){8-9}
{$n$} & {median} & {max} & {median} & {max} & {median} & {max} & {median} & {max}\\
\midrule
10 & 0.0140186585000208 & 0.0643094924999996 & 0.000366843500017922 & 0.000910092499992743 & 0.000519152999999939 & 0.00416178599998318 & 0.000819917500024303 & 0.00260278150000204
 \\
20 & 0.0813472499999932 & 0.361370027500016 & 0.000615681000027735 & 0.00161253850001231 & 0.00115587650000748 & 0.0118527379999591 & 0.00201029750002135 & 0.00734257650000814
 \\
30 & 0.207516140499933 & 0.943375459000208 & 0.00082982149979216 & 0.00287940899988826 & 0.00331893549991946 & 0.0264057159999993 & 0.00365473599993038 & 0.0145516520000228
 \\
40 & 0.4695303255002725 & 2.11221717900025 & 0.0010446840003623952 & 0.00216632799993022 & 0.0068672602499191306 & 0.0338505125000665 & 0.0057162397498586905 & 0.0203339299996514
 \\[0.5em]
60 &  &  & 0.00139869899976475 & 0.00291832249968138 & 0.01342819525007145 & 0.0566913724996994 & 0.009515700000520155 & 0.0306251470001371
 \\
80 &  &  & 0.0017564829995535551 & 0.00344350449995545 & 0.0240553282501423 & 0.100308998500623 & 0.0166582215001654 & 0.0613255340003889
 \\
100 &  &  & 0.00205603250105924 & 0.00403033249949658 & 0.0334281819996249 & 0.150278575499215 & 0.0225599480008896 & 0.0681703480004217
 \\
125 &  &  & 0.00250071300069976 & 0.00537218249974103 & 0.046975634500086 & 0.217761577498095 & 0.0339537715008191 & 0.101314634001028
 \\[0.5em]
150 &  &  & 0.00288767725032812 & 0.00553443600074388 & 0.06150543725016175 & 0.232505231999312 & 0.041352329749315644 & 0.133887884499927
 \\
200 &  &  & 0.00357625499964342 & 0.00723971800107392 & 0.100988728998345 & 0.354232027995749 & 0.0645319550003478 & 0.193955706500674
 \\
250 &  &  & 0.004320714000641599 & 0.00853678250132361 & 0.1378294190017185 & 0.458281022998563 & 0.0861407329994108 & 0.260131919498235
 \\
300 &  &  & 0.0050726427580229955 & 0.00968493349955679 & 0.17374915100299398 & 0.73996013499891 & 0.11024918075054299 & 0.407215889999861
 \\[0.5em]
350 &  &  & 0.00557878524705302 & 0.0105967549898196 & 0.20025592750243948 & 0.80860402899998 & 0.1284253874982825 & 0.457386415000656
 \\
400 &  &  & 0.0063722802442498505 & 0.0117612724998253 & 0.2493986304980355 & 0.887697580499662 & 0.175069794000592 & 0.589235497987829
 \\
450 &  &  & 0.00670990499202162 & 0.0135369370000262 & 0.282496769010322 & 1.09625077999954 & 0.175411713018548 & 0.58701009500146
 \\
500 &  &  & 0.00746870102011599 & 0.0150049549993128 & 0.38368185251602 & 1.54150053602643 & 0.256470213498687 & 0.74614374700468
 \\[0.5em]
550 &  &  & 0.008528186997864395 & 0.0161897120124195 & 0.4155791442508415 & 1.66693828100324 & 0.2613059092182085 & 0.859130185999675
 \\
600 &  &  & 0.00889877423469443 & 0.0171267999685369 & 0.48615729904122396 & 1.66015765299744 & 0.3072608275397215 & 0.969043154000246
 \\
650 &  &  & 0.00935010047396645 & 0.0189871419570409 & 0.520370340003865 & 1.70765144302277 & 0.320546510512941 & 0.991329870501431
 \\
700 &  &  & 0.0100329295382835 & 0.021339038037695 & 0.548667960450985 & 2.01827158900051 & 0.323925908945967 & 1.18718045200148
 \\[0.5em]
800 &  &  & 0.01158255774998905 & 0.0219055579999576 & 0.7296381919999535 & 2.50292197999079 & 0.452031759022134 & 1.526597103104
 \\
900 &  &  & 0.0129740017500808 & 0.0294806054998844 & 0.8663535162500009 & 3.218502634 & 0.5369776137499684 & 1.86649709700032
 \\
1000 &  &  & 0.01420388849976465 & 0.0297695079998448 & 1.019634877000045 & 4.07081257299978 & 0.6237622695000485 & 1.96960285400019
 \\
1100 &  &  & 0.0151214372499453 & 0.0321602915000767 & 1.23957746949975 & 3.84622054600004 & 0.717238796499942 & 2.04014411499975
 \\[0.5em]
1200 &  &  & 0.016400604250293348 & 0.0358420445002139 & 1.251472233499955 & 4.01051462799933 & 0.7337674704999699 & 2.28263964300004
 \\
1300 &  &  & 0.01835746649999235 & 0.0392718209996019 & 1.50691839799947 & 5.62713758700011 & 1.00147388900041 & 3.09502752200024
 \\
1400 &  &  & 0.0193541612502486 & 0.0438862330011034 & 1.64114956399999 & 5.70777366099901 & 1.02624043200012 & 3.41523584700008
 \\
1500 &  &  & 0.0212793032501395 & 0.0430127745000846 & 1.99008823950044 & 7.97812436299864 & 1.26153877600154 & 3.94102434500019
 \\[0.5em]
2000 &  &  & 0.026626784999734802 & 0.0546532909993402 & 2.8216841044999947 & 9.1515475010001 & 1.731934649499635 & 4.85433838099925
 \\
2500 &  &  & 0.0328613012512733 & 0.0707595024998682 & 3.53365639799995 & 12.0036069259986 & 2.187277281499855 & 6.84082411999771
 \\
3000 &  &  & 0.0391879192511624 & 0.0817730404996837 & 5.616593058501165 & 14.9497199569996 & 3.159590347499825 & 9.75131499200143
 \\
\bottomrule\end{tabular}

\end{table}

\end{document}